\documentclass[%
 aip,
 amsmath,amssymb,
 reprint,%
]{revtex4-2}

\usepackage{graphicx}
\usepackage{dcolumn}
\usepackage{bm}

\usepackage[utf8]{inputenc}
\usepackage[T1]{fontenc}
\usepackage{mathptmx}
\usepackage{braket}             
\usepackage{bm}                 
\usepackage{stackengine,graphicx}
\usepackage[caption=false,font=small]{subfig}
\begin{document}

\preprint{AIP/123-QED}

\title[Molecular reorganization energy in quantum-dot cellular automata switching] {Molecular reorganization energy in quantum-dot cellular automata switching}

\author{Craig S. Lent}%
 \email{lent@nd.edu}
 \author{Subhash S. Pidaparthi}
\affiliation{ Department of Electrical Engineering, University of Notre Dame, Notre Dame, IN 46556, USA 
}%


\date{\today}

\begin{abstract}
 We examine the impact of the intrinsic molecular reorganization energy on switching in two-state quantum-dot cellular automata (QCA) cells. Switching a bit involves an electron transferring between charge centers within the molecule. This in turn causes the other atoms in the molecule to rearrange their positions in response. We capture this in a model that treats the electron motion quantum-mechanically, but the motion of nuclei semiclassically. This results in a non-linear Hamiltonian for the electron system.  Interaction with a thermal environment is included by solving the Lindblad equation for the time-dependent density matrix. The calculated response of a molecule to the local electric field shows hysteresis during switching when the  sweep direction is reversed. The relaxation of neighboring nuclei increases localization of the electron, which provides an intrinsic source of enhanced bistability and single-molecule memory.  This comes at the cost of increased power dissipation.
\end{abstract}

\maketitle

%

\section{Introduction}
Quantum-dot cellular automata (QCA) is a  potential alternative to traditional complimentary metal-oxide semiconductor (CMOS) transistor logic for nanoelectronics. QCA uses the configuration of localized charge to represent binary information. \cite{LentTougawPorodBernstein:1993, TOUGAW1994, lent1997device, LentTougawArchitecture, WOS:000321668000004, WOS:000461570300005,  WOS:000380026400002, WOS:000590162700054,  WOS:000381320200013} A single mobile electron can be localized in a  quantum dot, which is simply region of space with a surrounding potential barrier that quantizes the charge enclosed. The arrangement of charge among multiple dots in a QCA cell encodes information and electron transfer (ET) between dots through quantum mechanical tunneling  enables switching. The electric field from one cell influences  neighboring cells enabling device operation. A two-dot cell captures the most basic elements of QCA operation. Three-dot or six-dot cells can support clocked control of the flow of information, and allow power gain by replacing energy lost to dissipative processes through work done by the clock. 

The QCA approach has been implemented in small metallic dots.\cite{LENT1994} Logic gates including majority gates,\cite{Amlani1999a} digital latches,\cite{Orlov2001} and shift registers\cite{Amlani1998, Kummamuru2003} have been demonstrated. Power gain\cite{Kummamuru2002} and signal restoration have also been demonstrated in these systems.   

Semiconductor QCA cells have been demonstrated in GaAs\cite{GaAlAsQCA} and silicon, \cite{SiliconQCAKern2003} using dots formed by surface depletion gates.
Another means of forming QCA cells in silicon is using a few implanted donor atoms to form the dots.  \cite{MiticSiDonorQCA2006} Molecular-scale QCA operation at room temperature has been achieved using dots formed by STM patterning of single dangling bonds on a hydrogen-passivated silicon surface. \cite{haider2009controlled}

Achieving QCA operation using single molecules\cite{lent2000bypassing,lieberman2002quantum, Lent2003} has focused attention on the richly-studied field of mixed-valence chemistry. In these molecules, the role of dots is played by charge centers, frequently formed by metal atoms surrounded by coordinating ligands. The central metal atom can exist in more than one charge state and can be reversibly oxidized or reduced without breaking any chemical bonds. A second charge center within the molecule is connected through a bridging ligand so an electron can tunnel between dots. Of interest for QCA operation is the case when the charge centers are identical. If the bridging ligand presents a sufficient barrier to tunneling, the mobile charge will localize on one of the dots. The localization can be enhanced by the relaxation of the other atoms in molecule in response to the presence or absence of the charge on one or the other dot. 

\begin{figure}[t]
\centering
\includegraphics[scale=0.4]{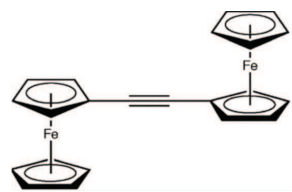}
\caption{Chemical structure of the mixed-valence diferrocenylacetylene (DFA) molecule. The iron centers in the two ferrocenyl groups act as dots for charge localization, forming a molecular double-dot. An  extra electron present on one or the other dot encodes binary information. The energy associated with the relaxation of each  ferrocenyl group in response to the presence or absence of the electron is the reorganization energy.}
\label{fig:DFA}
\end{figure}

A simple example of such a mixed-valence molecule is the diferrocenylacetylene (DFA) \cite{kramer1980electron} molecule shown in Figure \ref{fig:DFA}. In DFA, a pair of ferrocenyl groups are connected through a bridging ligand. The two iron centers act as the two dots where an electron can be localized. 

\begin{figure}[t]
\centering
\includegraphics[scale=0.35]{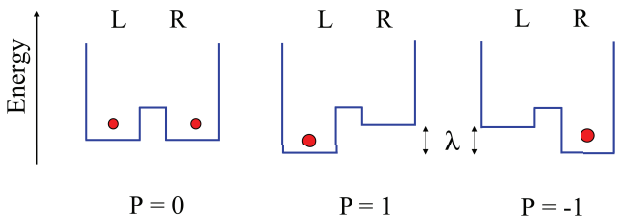}
\caption{Schematic representation of effect of reorganization energy $\lambda$. A molecular double-dot can be viewed as a double potential well holding one mobile electron. If the electron is on the left (right) dot the polarization is +1 (-1).  Electron occupation of a dot induces a relaxation of the positions of the surrounding atoms in response. This relaxation lowers the energy of the occupied dot compared to the unoccupied dot by the reorganization energy $\lambda$.}
\label{fig:reorganization}
\end{figure}

The potential advantages of single-molecule devices of this type are several. The functional density could be enormous because the footprint of each molecule on the surface is of the order of a square nanometer. In reality, it would not be practical to separately address and control each molecule separately, so  clocking schemes and appropriate computational architectures, such as a Kogge network,\cite{KoggeNetworks2016} would be necessary. Another advantage is that the intrinsic speed of such devices could be very high. Electron transfer in mixed-valence molecules can occur at  picosecond timescales. \cite{KubiakRapidTransfer1997}  Finally, very low power dissipation may be possible because switching the bit  requires only the motion of a single electron within the molecule. In contrast to molecular diodes or transistors, no current flow through the molecule is necessary. 

Several candidate molecules for QCA have been synthesized and characterized. Fehlner {\em et al.}  have synthesized QCA double-dots which were then covalently bonded to a semiconductor substrate.\cite{Qi2003,2003JACS_Fehlner_BuildingBlocks}  They demonstrated controlled switching of the molecular charge between dots using external electrodes. The localization of charge in mixed-valence molecules has been imaged using STM. \cite{2013DFA, Lu10,  WasioSTM2012, QuardokusThroughBond2012} 

Most mixed-valence species are ions. This may present complications due to the presence of neutralizing counterions that could bias the charge configuration of the QCA molecule. One strategy is to include an electron donor (or acceptor) at a symmetric position within the molecule. This creates a neutral mixed-valence zwitterion. \cite{LuLentCounterionFree2013, LuLentSelfDoping2011, HaletZwitterion2020}  Christie {\em et al.} have synthesized such a molecule, which could form the basis of a clockable three-dot QCA cell.\cite{christie2015synthesis}

Molecular QCA operation depends fundamentally on electric field driven ET between two dots within the molecule. We focus here on a simple model system that captures the essential features of this operation. Electron transfer in these systems is driven by an electric field, either produced by neighboring QCA molecules, or by clocking electrodes, or by input electrodes at the edges of a circuit array. This is in contrast to the usual situation in mixed valence chemistry, where ET is produced by either the random motion of solvent molecules in solution or by incident light. 

Full {\em ab initio} quantum chemistry calculations of molecular states are possible for the static problem in an electric field, \cite{Lu2008, IsaksenLentMolecular_QCA2003, lent2003clocked, WOS:000380026400002, LuLentDynamics2007} but calculation of time-dependent switching dynamics are too costly and one must use reduced-state models that capture the relevant physics. The parameters of the simplified system can be taken from either quantum chemistry calculations or from experiments. 

The key parameters for such a mixed-valence double dot system model are: the inter-dot Hamiltonian tunneling matrix element, here called $\gamma$ and often called $H_{ab}$ in the chemical literature; the on-site energy difference due to the applied field, here called $\Delta$; the Marcus reorganization energy $\lambda$ that captures the effect of the nuclear relaxation due to charge occupancy of one or the other dot; the energy relaxation time $T_d$ which characterizes the energetic coupling between the molecule and the surrounding environment; and the environmental thermal energy $k_BT$.  

The inter-dot tunneling energy $\gamma$ determines the intrinsic time scale for the system and can be varied over 12 orders of magnitude  by adjusting the bridging ligand. \cite{MarcusNobelRMP1993, KubiakRapidTransfer1997}  Reference \citenum{QuardokusThroughBond2012} shows the  dramatic effect on localization of simply altering the geometry of the connection between the dots and the bridging cyclopentadiene ring from the {\em meta} to {\em para} configuration. 

A molecular double-dot can be simply modelled as a double potential well along the spatial axis connecting the two metal centers.  (Note that this is not the Marcus double well as a function of the reaction coordinate). The reorganization energy $\lambda$ lowers the energy of the occupied well compared to the unoccupied well due to the relaxation of  other atoms, either those in the molecule itself or in the surrounding environment, as shown schematically in Figure \ref{fig:reorganization}. The value of $\lambda$ in solution can be affected by changing the solvent or by altering the chemical structure of the dot ligands. The reorganization energy  can also be strongly affected by the attachment of the molecule to a surface. If the effective stiffness of the dot ligands is increased because of the constraint of the surface, the reorganization energy becomes smaller. (For a given force, a spring with a larger spring constant will store less energy than one with a smaller spring constant).  A recent examination of the reorganization energy of a molecule with two ferrocene redox centers connected by a naphthalene linker and deposited on a NaCl substrate showed a reorganization energy in the few meV range, rather than the few hundred meV range for ferrocene in solution. \cite{BergerQuantumSTMDoubleDot2020, PaulTunableReorg2019} In a similar way, the strength of the coupling to the thermal environment which determines $T_d$ is dependent on the details of the surface attachment. 

The reorganization energy $\lambda$ can be obtained theoretically by high quality quantum chemistry calculations.  The nuclear motion of all the atoms in the molecule that are coupled to the electron transfer can be expressed in terms of the normal modes of vibration. These can then be treated quantum mechanically, though again this is computationally costly because of the need to account for the harmonic oscillator degrees of freedom. \cite{BittnerET2014, Blair2016}

We employ a parameterized Hamiltonian to describe the electron moving between the two charge centers under a driving field. The electron motion is linearly coupled to the nuclear reorganization via a single vibrational mode (the antisymmetric breathing mode). The energy transferred to the nuclear degrees of freedom is accounted for semiclassically so energy is conserved in the electron-vibron system. This leads to a nonlinear Hamiltonian for the electron degrees of freedom. Finally, the interaction with the thermal environment is included through Lindblad operators.  Our focus is on accounting for the ``self-trapping'' aspect of the relaxation of nuclear coordinates associated with ET, the associated hysteresis in switching behavior,  and the effect  on power dissipation.

Low power dissipation for device switching is a key requirement for any nanoelectronics. For classical systems, the dissipated energy exhibits an inverse linear dependence on switching time ($1/T_s$). To calculate the energy dissipated for a quantum system, a first-order approach is to consider  an isolated system and calculate the residual energy that remains in the system at the end of a switching event. This is the energy that will need to be eventually dissipated. For quantum switching, there is a remarkable  exponential reduction\cite{pidaparthi2018exponentially} in the residual  energy with switching time: $E\propto e^{-a T_s}$.   For a rigid molecule, this has been extended  to calculate dynamically the flow of  energy between the molecule and the environment. For rapid switching, the power dissipation shows the characteristic quantum exponential decrease as the switching time is increased.  For intermediate switching speeds there is an inverted region in which power dissipation actually increases with slower switching speed because of excitation from the thermal environment. At large switching times the power dissipation follows the classical $1/T_s$ dependence.\cite{pidaparthi2021energy}

Here we extend the model to include the effect of atomic reorganization. Section II describes the mathematical model for the system and interaction with the environment. Section III shows the impact of reorganization energy on polarization and switching dynamics. We solve the switching dynamically using the Lindblad equation and observe hysteresis in cell polarization when the bias sweep direction changes. Thus we demonstrate that a single mixed-valence molecule can function as a memory element.  In Section IV we calculate the power dissipated by switching a bit.  A higher reorganization energy causes an increase in  dissipated power. 

\section{Two-State Model}

\begin{figure}
\centering
\includegraphics[scale=0.5]{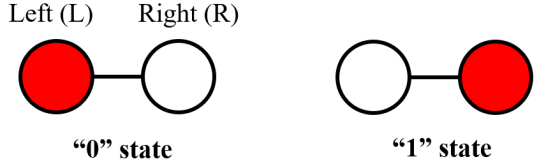}
\caption{Schematic of a two-state quantum-dot system. The black circles represent the quantum dots, the lines connecting the dots represent the inter-dot tunneling paths, and the red circles denote the localized charge. When charge is localized on the left dot (L), the system is in quantum state $\Ket{L}$ and the polarization $P=1$ representing a binary 0. When the charge is localized on the right dot (R), the system is in quantum state $\Ket{R}$ and the polarization $P=-1$ representing a binary 1.}
\label{fig:twodotsystem}
\end{figure}

A two-state quantum-dot system is illustrated schematically in Figure \ref{fig:twodotsystem}.  It is the simplest QCA element consisting of two dots and a mobile charge. Charge transport between dots is through quantum-mechanical tunneling. We use $\Ket{L}$ and $\Ket{R}$ as the basis states and the binary digital data in the two-state QCA is encoded in the polarization $P$. When the charge is fully localized on the left dot, the system is in state $\Ket{L}$ and $P=1$ representing a binary 0. When the charge is fully localized on the right dot, the system is in state $\Ket{R}$ and $P=-1$ representing a binary 1. 

\begin{figure}
\centering
\includegraphics[scale=0.4]{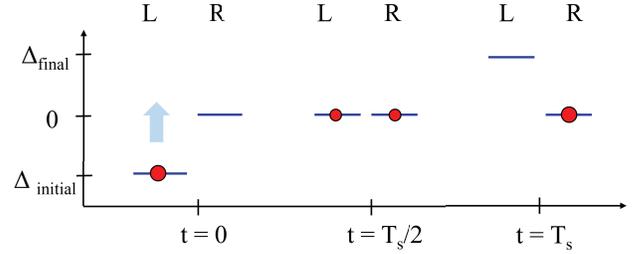}
\caption{Schematic representation of charge transfer during switching. At $t=0$ system is in ground state and charge is localized on left dot L. Bias $\Delta(t)$ is applied to  the left dot and is linearly increased from $\Delta_{initial}$ to $\Delta_{final}$ over the switching time $T_s$. Midway through the switching event at crossover $t=T_s/2$, the energies of the two dots are matched. At the end of the switching event $t=T_s$, charge is localized on the right dot and the electron transfer is complete.}
\label{fig:switching}
\end{figure}

Charge is transferred by changing the relative bias between the two dots. A simple switching operation is shown in Figure \ref{fig:switching}. At $t=0$, the charge is localized on the left dot. A time varying bias $\Delta(t)$ is applied to the left dot and the right dot is held at 0. $\Delta(t)$ is linearly increased from $\Delta_{initial}$ to $\Delta_{final}$ over switching time $T_s$. Midway through the switching event at crossover $t=T_s/2$, the energies of the two dots are matched. At the end of the switching event $t=T_s$ if the bias $\Delta$ is large enough, the charge is almost entirely localized on the right dot and the electron transfer is complete. Polarization changes continuously from a polarization $P=1$ at $t=0$ to a polarization of $P=-1$ at $t=T_s$ in this example. We interpret polarization near 1 as a binary 0 and polarization near -1 as a binary 1. \\ 
    %
    %
The density operator $\hat{\rho}$ can be written in terms of Pauli operators as
\begin{equation}
    \hat{\rho}=\frac{1}{2}\Big(\hat{I} +  \left<\hat{\sigma}_x\right>\hat{\sigma}_x +  \left<\hat{\sigma}_y\right>\hat{\sigma}_y +  \left<\hat{\sigma}_z\right>\hat{\sigma}_z \Big)
\end{equation}
where
\begin{equation}
   \left<\hat{\sigma}_i\right> = \mathrm{Tr}\big(\hat{\rho}\hat{\sigma}_i\big).
\end{equation}
The density operator is Hermitian with unit trace and the vector $\left( \left<\hat{\sigma}_x\right>,  \left<\hat{\sigma}_y\right>, \left<\hat{\sigma}_z\right>\right)$ will  lie on or within a unit sphere (commonly called the Bloch sphere):
\begin{equation}\label{eq:Bloch}
    \sqrt{|\left<\hat{\sigma}_x\right>|^2 + |\left<\hat{\sigma}_y\right>|^2 + |\left<\hat{\sigma}_z\right>|^2} \leqslant 1.
\end{equation}
\indent The Hamiltonian operator for the simplified electronic two-state subsystem is written as
\begin{equation}
    \hat{H}_E(t) = -\gamma\hat{\sigma}_x + \frac{\Delta(t)}{2}\big(\hat{\sigma}_z+ \hat{I}\big).
\end{equation}
\label{eq:Hamiltonian1}
\noindent The characteristic time scale for dynamics is 
\begin{equation}
T_\gamma=\frac{\pi\hbar}{\gamma}.   \label{eq:taudef}
\end{equation}
\indent The natural scales for the problem are to measure energy in units of $\gamma$ and time in units of $T_\gamma$.\\
The polarization $P$ is the expectation value of $\hat{\sigma}_z$
\begin{equation}\label{eq:Polarization}
    P = \left<\hat{\sigma}_z\right>= \mathrm{Tr}\big(\hat{\rho}\hat{\sigma}_z\big).
\end{equation}

When the electronic charge occupies a dot, the atoms around the dot shift their positions in response. This lowers the total energy for the electron being on that dot. The energy associated with this relaxation of nuclear positions is the  reorganization energy of Marcus theory $\lambda$. \cite{marcus1956electrostatic, marcus1956theory} The relaxation is modeled as the lowering of the energy of the dot when occupied by the electron (as in Figure \ref{fig:reorganization}).

The Hamiltonian for linear coupling\cite{PKS1978JAmChemSoc} between the  electron and the motion of the surrounding ligands can be written in terms of $\lambda$:
\begin{equation}\label{eq:HSL}
    \hat{H}_{EL} = -\frac{\lambda}{2}\hat{\sigma}_z\left<\hat{\sigma}_z\right>.
\end{equation}
\noindent The anti-symmetric oscillating motion of the left and right dot ligands we treat as a classical harmonic driven by the imbalance between the electron occupancy between the left and right dots.  The Hamiltonian operator representing this  energy stored in the displaced ligand positions (classically $\frac{1}{2}k x^2$) is written as
\begin{equation}\label{eq:HL}
     \hat{H}_L = \frac{1}{2}\bigg(\frac{\lambda}{2}\bigg)\left<\hat{\sigma}_z\right>^2 \hat{I}.
\end{equation}
\noindent Accounting in this way for energy as it moves back and forth between the electronic and vibrational degrees of freedom guarantees conservation of total energy.

The  Hamiltonian is the sum of the Hamiltonians of the electronic subsystem  $\hat{H}_E$,  the electron-ligand interaction $\hat{H}_{EL}$, and the ligand subsystem $\hat{H}_L$
\begin{equation}\label{eq:Hamiltonian2}
    \hat{H} = \hat{H}_E + \hat{H}_{EL} + \hat{H}_L.
\end{equation}
\begin{equation}
\label{eq:Hamiltonian}
\hat{H} = -\gamma\hat{\sigma}_x + \frac{\Delta}{2}\big(\hat{\sigma}_z+ \hat{I}\big) -\frac{\lambda}{2}\hat{\sigma}_z\left<\hat{\sigma}_z\right> + \frac{1}{2}\bigg(\frac{\lambda}{2}\bigg)\left<\hat{\sigma}_z\right>^2 \hat{I}.
\end{equation}
The Hamiltonian depends on the electron state through the expectation value $\braket{\hat{\sigma}_z}$. This non-linearity is fundamentally due to the fact that  the effect of the ligand distortion on the electron system has been treated semiclassically.

When a system is in equilibrium with an environment at temperature $T$ the steady state density operator $\hat{\rho}_{ss}$ can be written as
\begin{equation}\label{eq:rho_ss}
\hat{\rho}_{ss}= \frac{e^{\frac{-\hat{H}(\hat{\rho}_{ss})}{k_BT}}}{\mathrm{Tr}\Big(e^{\frac{-\hat{H}(\hat{\rho}_{ss})}{k_BT}}\Big)}.
\end{equation}
The steady state Hamiltonian $\hat{H}_{ss}=\hat{H}(\hat{\rho}_{ss})$ depends on the steady state density operator $\hat{\rho}_{ss}$  through Eq.  (\ref{eq:Hamiltonian}). For given values of $\gamma,\; \Delta$ (which may vary in time), $\lambda$, and $T$, Eq. (\ref{eq:rho_ss}) must be solved self-consistently.  We will see that in some circumstances multiple  solutions are possible for the same parameters.

\indent For an isolated system, the time evolution of the density operator $\hat{\rho}$ is unitary and can be solved using quantum Liouville equation:
\begin{equation}
\frac{\partial \hat{\rho}}{\partial t} = \frac{1}{i\hbar}\big[\hat{H}(t), \hat{\rho}\big].
\label{eq:vonNeumann}
\end{equation}
\noindent An open system interacts with the environment and the time evolution of the density operator is non-unitary. For Markovian systems, under  reasonable assumptions, the density operator can be solved using a Lindblad equation: \cite{breuer2002theory,lindblad1976generators, gorini1976completely}
\begin{equation}
\label{eq:LindbladEquation}
\frac{\partial \hat{\rho}}{\partial t}= \frac{1}{i\hbar}\big[\hat{H}(t),\hat{\rho}\big] + \underbrace{\sum_k\Big(\hat{L}_k\hat{\rho}\hat{L}_k^{\dagger}-\frac{1}{2}\big\{\hat{L}_k^{\dagger}\hat{L}_k,\hat{\rho}\big\}\Big)}_{\textstyle \mathbb{D}}
\end{equation}
where $\{\hat{A},\hat{B}\}$ is the anti-commutator of operators $\hat{A}$ and $\hat{B}$. The first term in Eq. (\ref{eq:LindbladEquation}) governs the unitary evolution of the density operator and is identical to Eq. (\ref{eq:vonNeumann}). The Lindblad dissipator $\mathbb{D}$ models the system-environment interaction including dissipation and decoherence. The $\hat{L}_k$ are the  Lindblad operators. We model the coupling of the system to the thermal environment for our two-state system as:
\begin{equation}
   \hat{L}_1(t)=\sqrt{\frac{1}{T_d}}\bigg[\ket{u_1(t)}\bra{u_2(t)}\bigg]
   \label{eq:LindbladOperator1}
\end{equation}
\begin{equation}
     \hat{L}_2(t)=\sqrt{\frac{1}{T_d}}
     \bigg[e^{-\frac{E_2(t)-E_1(t)}{2k_BT}}\ket{u_2(t)}\bra{u_1(t)}\bigg].
    \label{eq:LindbladOperator2}
\end{equation}
Here $\hat{H}(t)\ket{u_k(t)}=E_k(t)\ket{u_k(t)}$, $k_B$ is the Boltzmann constant, and $T$ is the temperature of the environment (bath). $T_d$ is a relaxation time or dissipation time and is a measure of the coupling strength between the system and the environment.  $T_d$ is small for a system coupled strongly to the thermal bath. For an isolated system with no interaction with the environment $T_d=\infty$ and both  $\hat{L}_1$ and $\hat{L}_2$ are zero.

The Lindblad operator $\hat{L}_1$ in Eq. (\ref{eq:LindbladOperator1}) models the decay of the system from the instantaneous excited state to the instantaneous ground state by dissipating energy to the environment. The Lindblad operator $\hat{L}_2$ in Eq. (\ref{eq:LindbladOperator2}) models the thermal excitation of the system from the ground state to the excited state by absorbing thermal energy from the environment. The operators are constructed such that when the potential driver stops at $t=T_s$,  the density operator will relax with  characteristic time $T_d$ to the thermal equilibrium density operator in Eq. (\ref{eq:rho_ss}).

\section{Polarization of an Open Two-State System during Switching}

\subsection{Hysteresis in Polarization during Dynamic Switching
\label{sec:Hysteresis}}

We calculate the polarization during a switching operation of an open two-state system with a Hamiltonian defined in Eq. (\ref{eq:Hamiltonian}). We solve the Lindblad equation in Eq. (\ref{eq:LindbladEquation}) for the density operator $\hat{\rho}(t)$ as a function of time and the polarization $P$ is calculated directly from the density operator using Eq. (\ref{eq:Polarization}). The system is in contact with the environment at temperature $T$ and the coupling between the system and environment is characterized by the dissipation time $T_d$. Initially at $t=0$, the bias is $\Delta_{min}=-25\gamma$ and the system is in equilibrium with the environment. The initial dot occupancy probabilities are determined by the steady state density density operator from Eq. (\ref{eq:rho_ss}). The bias is linearly increased from $\Delta_{min}=-25\gamma$ at $t=0$ to $\Delta_{max}=+25\gamma$ over switching time $t=T_s$ which we define as a positive sweep. After the positive sweep, the bias is held at $\Delta_{max}$ for a time sufficiently long such that system can come to equilibrium with the environment. Then the bias is linearly decreased from $\Delta_{max}$ back down to $\Delta_{min}$ over the same switching time $T_s$ which we define as a negative sweep. We explore the effect of changing the reorganization energy $\lambda$, switching time $T_s$, dissipation time $T_d$ and environmental temperature $T$.

\begin{figure}[h!]
\centering
\includegraphics[scale=0.6]{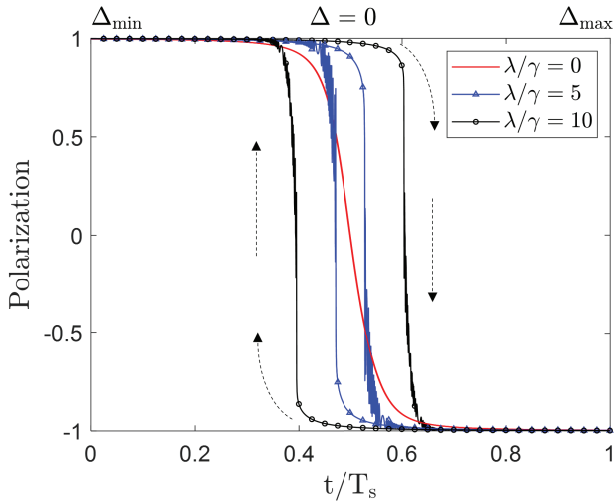}
\caption{Hysteresis in polarization and the impact of reorganization energy on an open system. Increasing reorganization energy $\lambda$ increases the hysteresis in polarization and causes switching to be less gradual. With sufficient $\lambda$, polarization is approximately equal to +1 or -1 at zero bias ($t=T_s/2$) depending on the bias sweep direction. In this calculation, the bias $\Delta$ linearly increases from $-25\gamma$ to $+25\gamma$ over switching time $T_s$ during the positive sweep and decreases from $+25\gamma$ to $-25\gamma$ over the same switching time $T_s$ during the negative sweep. Here $\lambda/\gamma=[0, 5, 10]$, $T_s=1000T_{\gamma}$, $T_d=10T_{\gamma}$ and  $k_BT=1\gamma$.}
\label{fig:Polarization1}
\end{figure}

Figure \ref{fig:Polarization1} shows the effect of the reorganization energy $\lambda$ for quasi-adiabatic switching ($T_s/T_{\gamma} \gg 1$). When there is no reorganization energy ($\lambda=0$), the polarization of the positive sweep and the negative sweep overlap. At zero bias ($\Delta$=0), the polarization is zero and therefore no binary information is stored in the two-state system. 

When the reorganization energy is introduced, the system exhibits hysteresis. The polarization is different depending on the sweep direction. The hysteresis widens with increasing reorganization energy. At $t=T_s/2$ when the bias is zero ($\Delta=0$), the polarization $P\approx1$ or $P\approx-1$ depending on the sweep direction. This means the binary information ("0" bit or "1" bit) is stored even when the bias is removed. The polarization is closer to 1 or -1 when the reorganization energy increases indicating bit storage is easier at higher reorganization energies. We will discuss this more in the next subsection. The slope of the polarization  becomes more abrupt due to the effect of the reorganization energy.

\begin{figure}[h!]
\centering
\includegraphics[scale=0.6]{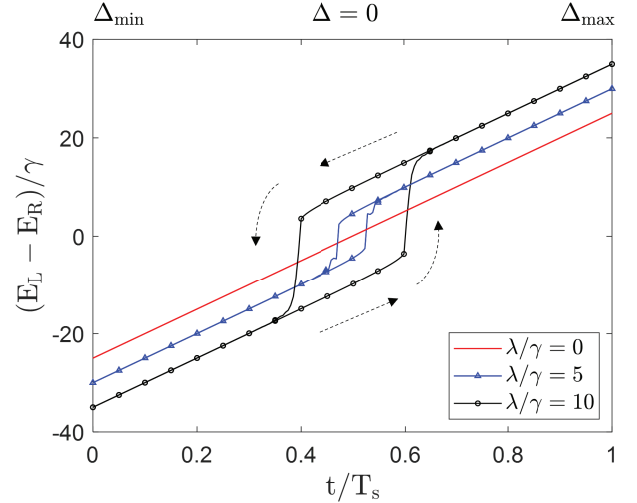}
\caption{Difference in onsite energy of dots $E_L-E_R$ for the switching operation shown in Figure \ref{fig:Polarization1}. With no reorganization energy $\lambda=0$, $E_L-E_R$ matches the applied bias $\Delta$. With a non-zero reorganization energy, the change in $E_L-E_R$  is faster (steeper slope) compared to the change in the applied bias. This causes the switching to be more abrupt and therefore less adiabatic.}
\label{fig:OnsiteEnergy}
\end{figure}
\indent The onsite energies of the left dot $E_L$ and right dot $E_R$ are defined as
\begin{eqnarray}
    E_L(t)=\bra{L}\hat{H}(t)\ket{L}\\
    E_R(t)=\bra{R}\hat{H}(t)\ket{R}
\end{eqnarray}

In Figure \ref{fig:OnsiteEnergy}, we plot the difference of the onsite energies of the left and right dots ($E_L-E_R$) for the quasi-adiabatic switching calculations shown in Figure \ref{fig:Polarization1}. When $\lambda=0$, $E_L-E_R$ simply matches the  applied bias. The difference linearly increases from $\Delta_{min}$ to $\Delta_{max}$ during the positive sweep and linearly decreases from $\Delta_{max}$ to $\Delta_{min}$ during the negative sweep.

\indent  Now consider the system with a nonzero reorganization energy. Initially when charge is localized on the left dot ($P=1$) the minimum value of $E_L-E_R$ is $\Delta_{min}-\lambda$. During the positive sweep as the bias increases, $E_L-E_R$ increases at the same rate. When $E_L$ comes closer to $E_R$, the charge transfer begins. Increasing charge occupation causes relaxation of the nuclei in the right dot and lowers the energy of right dot. The lowering of energy accelerates the charge transfer which further lowers the energy of the right dot. This phenomenon causes switching to be less adiabatic and more abrupt. This is evident from the steeper slope of $E_L-E_R$ in a system with reorganization energy compared to a system with $\lambda=0$. The polarization curves in Figure \ref{fig:Polarization1} become abrupt with increasing reorganization energy. At the end of the positive sweep, the charge is localized on the right dot ($P=-1$) and  $E_L-E_R$ is $\Delta_{max}+\lambda$. The same behavior is observed during the negative sweep but in the opposite direction. \\
\begin{figure}[h!]
\centering
\includegraphics[scale=0.6]{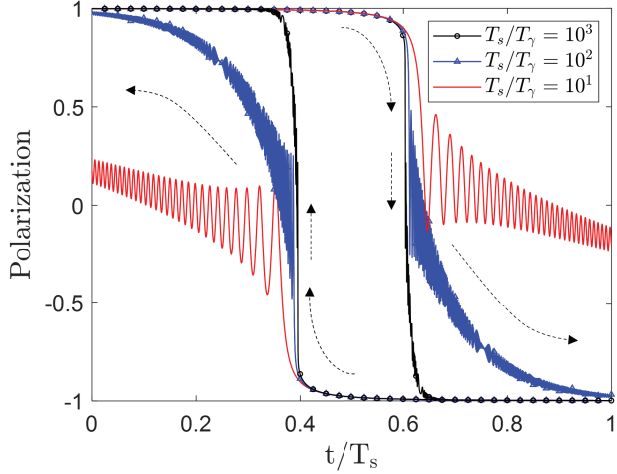}
\caption{Hysteresis in polarization and the impact of switching time for an open system. For quasi-adiabatic switching (large $T_s/T_\gamma$), charge transfer is smooth and complete. Switching too rapidly causes quantum oscillations and incomplete charge transfer. The  system may be left in excited state at the end of the switching event. In this calculation, the bias $\Delta$ linearly increases from $-25\gamma$ to $+25\gamma$ over switching time $T_s$ during the positive sweep and decreases from $+25\gamma$ to $-25\gamma$ over the same switching time $T_s$ during the negative sweep. Here  $T_s/T_{\gamma}=[10, 100, 1000]$, $\lambda=10\gamma$, $T_d=10T_{\gamma}$ and  $k_BT=1\gamma$. }
\label{fig:Polarization2}
\end{figure}
\indent Figure \ref{fig:Polarization2} shows the impact of switching time $T_s$.  When the switching is slow ($T_s/T_{\gamma}=1000$) and therefore quasi-adiabatic switching, the system follows the instantaneous ground state in accordance with adiabatic theorem. The charge switching is complete and the polarization changes from -1 to 1 and vice versa. When the switching is rapid ($T_s/T_{\gamma}$ is small), the charge transfer is not complete at end of the sweep ($t=T_s$) and we observe charge quantum oscillations between the two dots post-crossover. The system will return to ground state a sufficiently long time after the switching event.

\begin{figure}[h!]
\centering
\includegraphics[scale=0.6]{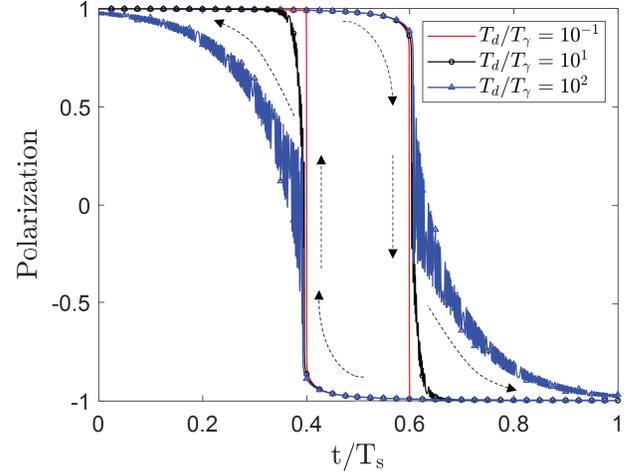}
\caption{Hysteresis in the polarization and the impact of coupling strength to the environment for an open system. Systems coupled strongly to environment (small $T_d$) reach steady state quickly. Weakly coupled systems (large $T_d$) take longer to reach steady state. In this calculation, the bias $\Delta$ linearly increases from $-25\gamma$ to $+25\gamma$ over switching time $T_s$ during the positive sweep and decreases from $+25\gamma$ to $-25\gamma$ over the same switching time $T_s$ during the negative sweep. Here $T_d/T_{\gamma}=[0.1, 10, 100]$, $T_s=1000T_{\gamma}$,  $\lambda=10\gamma$ and $k_BT=1\gamma$.}
\label{fig:Polarization3}
\end{figure}

Figure \ref{fig:Polarization3} shows the impact of coupling strength between the system and the environment.   When the system is strongly coupled to the environment ($T_d$ is small), any excess energy is immediately  dissipated and the switching is more abrupt. For systems that are weakly coupled to the environment ($T_d$ is large), the system takes longer to reach the ground state and the complete switching of the charge takes longer.

\begin{figure}[h!]
\centering
\includegraphics[scale=0.6]{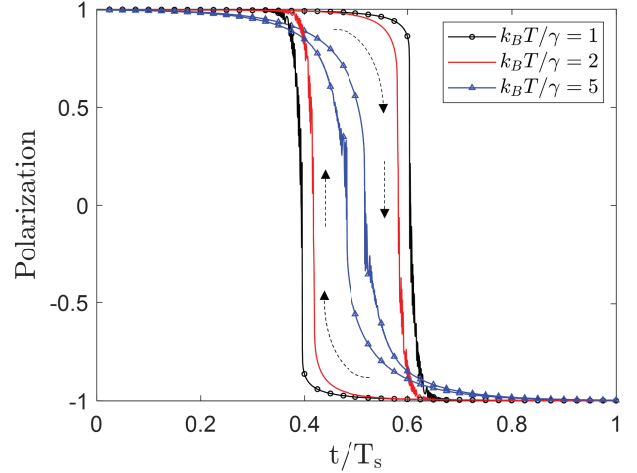}
\caption{Hysteresis in polarization and the impact of environmental temperature for an open system. Increasing temperature causes switching to be smoother and decreases hysteresis. In this calculation, the bias $\Delta$ linearly increases from $-25\gamma$ to $+25\gamma$ over switching time $T_s$ during the positive sweep and decreases from $+25\gamma$ to $-25\gamma$ over the same switching time $T_s$ during the negative sweep. Here $k_BT/\gamma=[1, 2, 5]$, $T_s=1000T_{\gamma}$, $T_d/T_{\gamma}=10$ and $\lambda=10\gamma$.}
\label{fig:Polarization4}
\end{figure}

Figure \ref{fig:Polarization4} shows the impact of environmental temperature. At higher environmental temperature, the hysteresis decreases and the switching becomes less abrupt. The system takes longer to reach the ground state due to  thermal excitation from the environment.

From Figures \ref{fig:Polarization1} and \ref{fig:Polarization4}, we observe that the width of the hysteresis is set by the reorganization energy $\lambda$ and environmental temperature $T$. Changing either the switching time $T_s$ or the dissipation time $T_d$ (in Figures \ref{fig:Polarization2} and \ref{fig:Polarization3}), does not change the hysteresis but only changes the time taken to reach steady state. 
\subsection{Memory Operations on a Two-state system with Reorganization energy}
\begin{figure}[h!]
\centering
\includegraphics[scale=0.49, trim={0.15cm 0 0.65cm 0},clip]{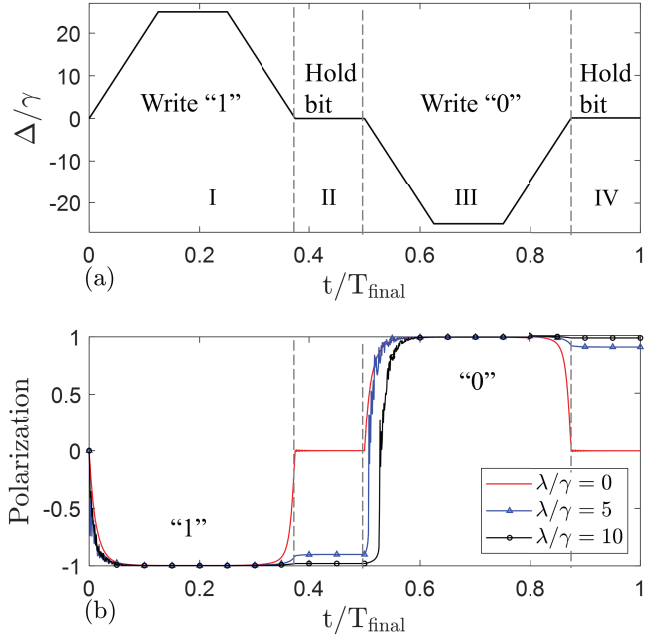}
\caption{Writing and holding 1 and 0 bits for different reorganization energy $\lambda$. In a system with sufficient reorganization energy, the polarization does not change significantly when the bias is removed. The system continues to store the bit (1 or 0) that is written. When $\lambda=0$, the polarization becomes zero when the bias is removed and the bit is not stored. Here $\lambda/\gamma=[0, 5, 10]$, $T_s=1000T_{\gamma}$, $T_d=10T_{\gamma}$ and  $k_BT=1\gamma$.}
\label{fig:OpenLindbladSwitch}
\end{figure}

The hysteresis observed in Figure \ref{fig:Polarization1} shows that a system with nonzero reorganization energy  can store a bit ("1" or "0") when the bias is zero during switching. We now perform a memory operation where we write and hold a "1" bit and then write and hold a "0" bit. The waveform of the applied bias is shown in Figure \ref{fig:OpenLindbladSwitch}a and the polarization response is shown in Figure \ref{fig:OpenLindbladSwitch}b.  We split this operation into four regions (I to IV) for the ease of explanation. Consider the polarization response when $\lambda=5\gamma$.

In the region labeled I, the "1" bit is written. Initially at $t=0$, the applied bias $\Delta=0$ and the system is in  equilibrium. The bias linearly increases and charge starts transferring. The bias is increased to $\Delta=+25\gamma$ by which point the charge is completely localized on the right dot with polarization $P=-1$. The system now represents a binary "1." The bias is then held constant for some time so the system can settle to a equilibrium. The bias is now linearly decreased from $\Delta=+25\gamma$ to $\Delta=0$.  In the system with significant reorganization energy ($\lambda=5\gamma$), the charge does not transfer back and the polarization continues to be approximately -1. The binary information written into the system is locked. The bit is stored even when the bias is removed because the nuclear relaxation has localized the charge on the right dot. 

In region labeled II, the bias is removed for a time duration of $10T_d$. The polarization for $\lambda=5\gamma$ does not change significantly and the binary information is not lost. The polarization continues to be approximately -1.

In the region labeled III, we write and hold a "0" bit and therefore the applied bias is of opposite sign of the waveform in region I. The bias is decreased from $\Delta=0$ to $\Delta=-25\gamma$ which causes the charge to localize on the left dot (P=1) and the system now represents a binary "0". When the bias is increased from $-25\gamma$ to back to 0, the polarization does not change significantly for $\lambda=5\gamma$. The polarization continues to be approximately equal to 1.

In the region labeled IV, the bias is held at 0 for a time duration of $10T_d$ similar to region II. When $\lambda=5\gamma$, the polarization continues to be approximately equal to 1 and the 0 bit is stored. 

Figure \ref{fig:OpenLindbladSwitch} also shows the impact of reorganization energy. When $\lambda=0$, there is no nuclear relaxation and the polarization is zero (bit not stored) in the absence of any applied bias. It is easier to hold binary information when the system has higher reorganization energy.

\subsection{Multiple Steady State Configurations}

In  \ref{sec:Hysteresis}, we calculated the polarization of a switching system dynamically using the Lindblad equation. We observed different values of polarization for the same value of applied bias. This means the system can be in more than one stable states. We now calculate the steady state solutions of the system using Eq. (\ref{eq:rho_ss}) to determine under what circumstances the system will support a single or multiple  equilibrium solutions.

We calculate the steady state density operator $\hat{\rho}_{ss}$ at a given bias by self-consistently solving Eq. (\ref{eq:rho_ss}) using the constraint in Eq. (\ref{eq:Bloch}). Without the loss of generality we can set $\left<\hat{\sigma}_y\right>=0$. To solve Eq. (\ref{eq:rho_ss}) iteratively, we start with an initial guess for the density operator and evaluate the RHS of Eq. (\ref{eq:rho_ss}) to obtain a value for $\hat{\rho}_{ss}$. This value is then substituted back into  (\ref{eq:rho_ss}) and another iteration is obtained. These iterations continue until we obtain a converged solution for the density operator that satisfies Eq. (\ref{eq:rho_ss}). We then calculate the polarization from the steady state density operator using Eq. (\ref{eq:Polarization}).

\begin{figure*}
\centering
\includegraphics[scale=0.6, trim={0.55cm 0 0.55cm 0},clip]{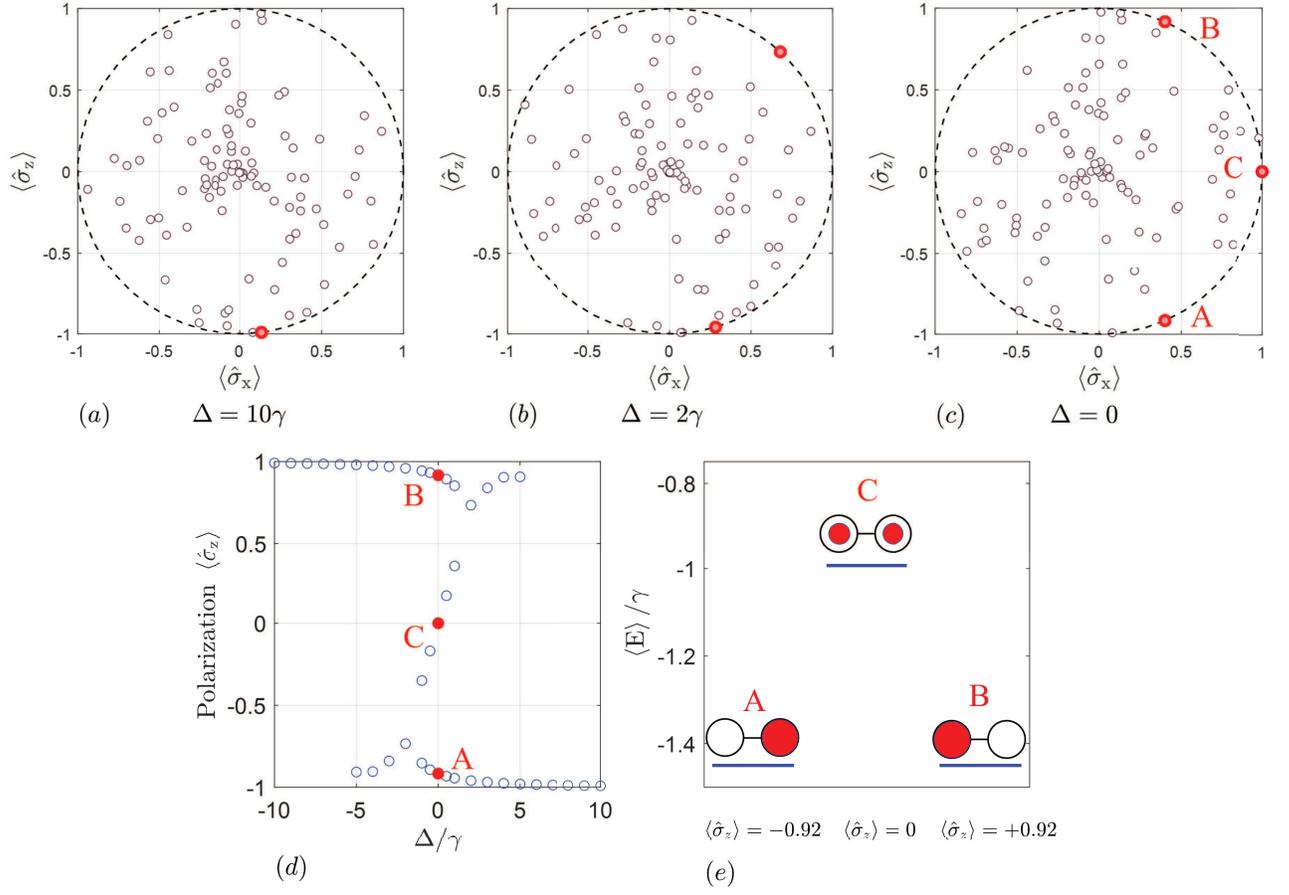}
\caption{Multiple equilibrium solutions of the density matrix for different applied biases. Solutions for the  steady-state density operator $\hat{\rho}_{ss}$ are obtained by randomly choosing an initial state and solving Eq. (\ref{eq:rho_ss}) self-consistently for $\lambda=5\gamma$ and $k_BT=0.25\gamma$.  In (a) to (c), the initial guesses are shown using open circle markers and steady-state solutions are show using filled circles. Steady-state solutions are calculated for three different bias values: (a) $\Delta=10\gamma$. All initial guesses converge to a single steady-state solution. (b) $\Delta=2\gamma$. There are two steady-state solutions and the system is bistable. (c) $\Delta=0$. There are three steady-state solutions which are labeled A, B and C. (d) Steady-state polarization as a function of the applied bias. For small bias values there are multiple steady-state solutions.  (e) Expectation value of energy $\left<E\right>$ for the three solutions A, B, and C at $\Delta=0$ shown in (c) and (d). $\left<E\right>$ for solutions A and B is lower than $\left<E\right>$ for solution C. The result is that solution C is not visited during dynamic switching. }
\label{fig:Random}
\end{figure*}

We calculate the steady state density operators for different values of bias $\Delta$. Figures \ref{fig:Random}a-c shows the results for bias $\Delta = 10\gamma$, $2\gamma$, $0$ respectively. Here we take reorganization energy $\lambda=5\gamma$ and environmental temperature $k_BT=0.25\gamma$. The initial guess for the density operator is randomly chosen to lie on or within the unit circle defined by $\left<\hat{\sigma}_x\right>$ and $\left<\hat{\sigma}_z\right>$. We then solve Eq. (\ref{eq:rho_ss}) iteratively to obtain a converged steady state solution for density operator and $\left<\hat{\sigma}_x\right>$ and $\left<\hat{\sigma}_z\right>$. We perform this calculation for 100 random initial guesses for each bias condition by solving Eq. (\ref{eq:rho_ss}) for each of the intial guesses. In Figure \ref{fig:Random}a, b, c the random initial guesses are shown using open circles and the converged solutions for each of these initial guesses are shown using filled circles.

Figure \ref{fig:Random}a shows the results when the applied bias is large ($\Delta = 10\gamma$). After solving Eq. (\ref{eq:rho_ss}), all the initial random states yield a single solution for $\hat{\rho}_{ss}$ with a polarization $\hat{\sigma}_z=-0.99$ and the charge is localized on the right dot.   
 
Figure \ref{fig:Random}b shows the results when the bias is lowered to $\Delta=2\gamma$. All the initial random states  converge to two possible solutions with polarization $\hat{\sigma}_z=-0.96$ and $0.73$. The two polarization solutions are of opposite signs but not symmetric across zero. At a low applied bias, the system is bistable. For these steady state solutions, the majority of the charge can be localized on either dot.

Figure \ref{fig:Random}c shows the results when $\Delta=0$. There are three possible steady states solutions (labeled A, B and C). Solutions A and B  with $\left<\hat{\sigma}_z\right>=-0.92$ and  $\left<\hat{\sigma}_z\right>=+0.92$ has the charge localized on the right or left dot respectively. Solution C with $\left<\hat{\sigma}_z\right>=0$ has the charge equally distributed between the dots. At a bias value close to 0, there can be three steady-state solutions.

Figure \ref{fig:Random}d  shows the steady state polarization as a function of bias $\Delta$.  Multi-state equilibrium solutions are observed when the bias is comparable to $\lambda + k_BT$. The polarization curve exhibits hysteresis. There are solutions near the origin (example solution C) that are not observed in the hysteresis behavior during dynamic switching calculations in Figure \ref{fig:Polarization1}. To understand why, we calculate the expectation value of energy for solutions A, B and C.  

In Figure \ref{fig:Random}e we calculate the expectation value of energy $\left<E\right>$ for the three solutions A, B, C at $\Delta=0$. $\left<E\right>$ for the two symmetric solutions A and B are equal and  $\left<E\right>=-1.45\gamma$. For solution C, $\left<\hat{\sigma}_z\right>=0$ and the expected value of energy $\left<E\right>=-0.99\gamma$. Solution C is not a stable state due to higher $\left<E\right>$ compared to solutions A and B. Solution C will not be visited during dynamic switching. The system will select either solution A or B depending on the sweep direction.

\section{Energy Dissipation due to Switching}

\subsection{Isolated System}
We can first calculate the energy dissipated by considering an isolated system, uncoupled to the environment, and calculate the excess energy $E_{excess}$ left in the system at the end of the switching event ($t=T_s$).  We know that for an open system this is the energy that will be eventually dissipated. What this approach ignores is dissipation, or thermal excitation, that occurs to the environment during the switching process.  For all calculations reported in this section,  $\Delta$ is linearly increased $-25\gamma$ to $+25\gamma$ over switching time $T_s$.  This is sufficient to strongly localize the charge both before and after switching.

Consider the switching event described in Figure \ref{fig:switching}. We define the excess energy of the system $E_{excess}$ as the difference between the expectation value of the energy $\left<E\right>$ and the ground state energy $E_1$ when the potential driver has stopped changing at $t=T_s$.
\begin{equation}\label{eq:Eexcess2dot}
    E_{excess} \equiv \left< E \right>(T_s)-E_1(T_s)  
\end{equation}

The dimensionless  adiabaticity parameter $\beta$ captures the effect of the parameters that impact the adiabaticity of the switching.

\begin{equation}
\beta \equiv \frac{2 \pi \gamma^2 T_s}{\hbar |\Delta_{final}-\Delta_{initial}|}
\label{eq:betaDef}
\end{equation}
For an isolated system with no reorganization energy $\lambda=0$, the excess energy $E_{excess}$ decreases \textit{exponentially} with $\beta$
\begin{equation}\label{eq:TwoStateEexcess}
E_{excess} \approx \Delta_{final}  e^{-\beta}.
\end{equation}
There is no fundamental lower limit to $E_{excess}$ as it can be decreased to any arbitrary low value by increasing the switching adiabaticity $\beta$. The adiabaticity $\beta$ can be made larger by switching more slowly (larger $T_s$), or using a smaller potential driver (smaller $\Delta$), or by increasing $\gamma$.

\begin{figure}
\centering
\includegraphics[scale=0.65]{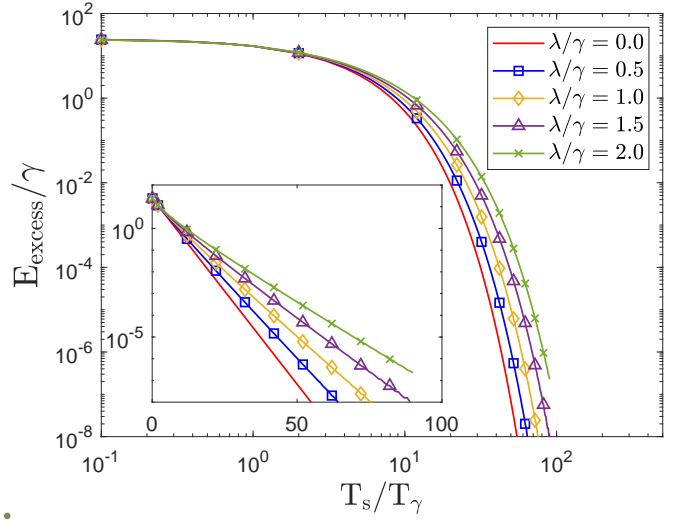}
\caption{Excess energy $E_{excess}$  at the end of a switching event for an isolated system as a function of the switching time $T_s$ for different values of the reorganization energy $\lambda$.  The inset is the same data with the horizontal axis plotted on a linear scale. $E_{excess}$ decreases exponentially with $T_s$ when $\lambda=0$. Increasing $\lambda$ causes switching to be somewhat less adiabatic and increases $E_{excess}$.}
\label{fig:TwoDotEexcess}
\end{figure}

The excess energy $E_{excess}$ left in the system at the end of the switching event is calculated using Eq. (\ref{eq:Eexcess2dot}) for different reorganization energies $\lambda$ and plotted as a function of switching time $T_s$ in Figure \ref{fig:TwoDotEexcess}. In the absence of reorganization energy $\lambda=0$, $E_{excess}$ decreases exponentially with increasing $T_s$  (and therefore $\beta$) as shown in Eq. (\ref{eq:TwoStateEexcess}). This result has been discussed previously. \cite{pidaparthi2018exponentially} The excess energy increases with the introduction of a nonzero reorganization energy $\lambda$. The reorganization energy causes the switching to be less adiabatic (see explanation for Figure \ref{fig:OnsiteEnergy} in the previous section). For the same switching speed, a system with higher reorganization energy has higher excess energy.

\subsection{System in contact with a thermal environment}

We now calculate $E_{diss}$, the total energy dissipated for a system in contact with a thermal environment due to a switching event. We first calculate the energy dissipated $E_{switch}$ {\em during} switching from $t=0$ to $t=T_s$. We do this by calculating the power flow and then integrating it over $t=0$ to $t=T_s$.  

\indent The total power flow $P_{total}$ into the system is the derivative of expected energy of the system:
\begin{equation}
P_{total}=\frac{\partial \left< E \right>}{\partial t}=
\frac{\partial }{\partial t} \mathrm{Tr}\big(\hat{\rho}\hat{H}\big)=\mathrm{Tr}\left(
\frac{\partial }{\partial t}\hat{\rho}\hat{H}\right)
\end{equation}
\begin{equation}\label{productrule}
P_{total} = \mathrm{Tr}\left(\frac{\partial \hat{\rho}}{\partial t}\hat{H} + \hat{\rho}\frac{\partial \hat{H}}{\partial t}\right)
\end{equation}
Using the Lindblad equation in Eq. (\ref{eq:LindbladEquation}), we can write
\begin{equation}\label{Pdissp}
P_{total}=\mathrm{Tr}\left(\Big(\frac{1}{i\hbar}[\hat{H},\hat{\rho}] + \mathbb{D}\Big)\hat{H}+ \hat{\rho}\frac{\partial \hat{H}}{\partial t}\right)
\end{equation}
\begin{equation}\label{eq:Pdissp1}
     P_{total}=\frac{1}{i\hbar}\mathrm{Tr}\left([\hat{H},\hat{\rho}]\hat{H}\right) + \mathrm{Tr}\left(\mathbb{D}\hat{H}\right) + \mathrm{Tr}\left(\hat{\rho}\frac{\partial \hat{H}}{\partial t}\right)
\end{equation}
The first term vanishes because of the cyclic property of the trace and Eq. (\ref{eq:Pdissp1}) reduces to
\begin{equation} \label{eq:powerflow2}
P_{total} = \mathrm{Tr}\big(\mathbb{D}\hat{H}\big) + \mathrm{Tr}\left(\hat{\rho}\frac{\partial \hat{H}}{\partial t}\right)
 \end{equation}
 By expanding the second term in the above equation using Eq. (\ref{eq:Hamiltonian2}) we obtain
 \begin{equation}\label{eq:powerflow3}
P_{total}=\underbrace{ \mathrm{Tr}\big(\mathbb{D}\hat{H}\big)}_{P_1} + \underbrace{\mathrm{Tr}\left(\hat{\rho}\frac{\partial \hat{H}_E}{\partial t}\right)}_{P_2} +\underbrace{\mathrm{Tr}\left(\hat{\rho}\frac{\partial \hat{H}_{EL}}{\partial t}\right)}_{P_3} + \underbrace{\mathrm{Tr}\left(\hat{\rho}\frac{\partial \hat{H}_L}{\partial t}\right)}_{P_4}
 \end{equation} 
 
 \noindent By substituting Eq. (\ref{eq:HSL}) in the third term in Eq. (\ref{eq:powerflow3}) we get
 
\begin{equation}
    P_3= \mathrm{Tr}\left(\hat{\rho}\frac{\partial \hat{H}_{EL}}{\partial t}\right)
    =-\frac{\lambda}{2} \mathrm{Tr}\left(\hat{\rho}\hat{\sigma}_z\frac{\partial\left<\hat{\sigma}_z\right>}{\partial t}\right)
\end{equation}
So
\begin{equation}
    P_3 =-\frac{\lambda}{2}\frac{\partial\left<\hat{\sigma}_z\right>}{\partial t}\left<\hat{\sigma}_z\right>
\end{equation}

 
 \noindent Similarly using Eq. (\ref{eq:HL}) we obtain
 
 \begin{equation}
   P_4=\frac{\lambda}{2} \left<\hat{\sigma_z} \right> \frac{\partial \left<\hat{\sigma_z} \right>}{\partial t}.
 \end{equation}
 
 \noindent The third term $P_3$ and fourth term $P_4$ in Eq. (\ref{eq:powerflow3}) cancel each other and therefore the total power flow into the system reduces to
 
 \begin{equation}\label{eq:powerflow4}
  P_{total} =\underbrace{ \mathrm{Tr}\left(\mathbb{D}\hat{H}\right)}_{P_1} + \underbrace{\mathrm{Tr}\left(\hat{\rho}\frac{\partial \hat{H}_E}{\partial t}\right)}_{P_2}
   \end{equation}
   
 \noindent We identify the second term $P_2$ in Eq. (\ref{eq:powerflow4}) as the power flowing into the system due to work done by the time-varying control electrodes during switching.
 
\begin{equation}
    P_{work}=\mathrm{Tr}\left(\hat{\rho}\frac{\partial \hat{H}_E}{\partial t}\right)
\end{equation}

\noindent We identify the first term $P_1$ in Eq. (\ref{eq:powerflow4}) as the instantaneous power flow \textit{into} the system from the environment. The power flowing \textit{from}  the system into the environment is obtained by simply changing the sign.

\begin{equation}\label{eq:powerdisp}
     P_{switch} = -\mathrm{Tr}\left(\mathbb{D}\hat{H}\right)
\end{equation}

\noindent The total energy dissipated \textit{during} the switching event $E_{switch}$ is the integral of instantaneous power dissipated $P_{switch}$  during entire switching event from $t=0$ to $T_s$.

\begin{equation}\label{eq:Edissswitch}
   E_{switch} =\int_{0}^{T_s} P_{switch}(t)dt=-\int_{0}^{T_s} \mathrm{Tr}\Big(\mathbb{D}\hat{H}\Big)dt
\end{equation}

\noindent We now consider the residual excess energy $E_{excess}$ left in the system when the potential driver stops at $t=T_s$. This residual excess energy will be dissipated after some time $t>T_s$. We define $E_{excess}$ as the difference between the non-equilibrium expectation value of the energy and the steady-state expectation value at the end of the switching event.

\begin{equation} \label{eq:Excess_ss}
   E_{excess} = \left< E \right>(T_s) - \left< E_{ss} \right>(T_s)
\end{equation}
\noindent where
\begin{eqnarray}
    \left< E \right>(T_s)&=& \mathrm{Tr}\Big( \hat{\rho}(T_s) \hat{H}(T_s)\Big)\\ 
    \label{eq:ExpEend}
    \left< E_{ss} \right>(T_s)&=& \mathrm{Tr}\Big(\hat{\rho}_{ss}(T_s)\hat{H}_{ss}(T_s)\Big) 
    \label{eq:ExpEqEnd}
\end{eqnarray}

\noindent The steady state values of $\hat{\rho}_{ss}$ and $\hat{H}_{ss}$ are obtained by solving Eq. (\ref{eq:rho_ss}) self consistently. The total energy dissipated $E_{diss}$ is the sum of the energy dissipated during switching $E_{switch}$ and excess energy  $E_{excess}$ left at the end of switching event which is eventually dissipated to the environment.

\begin{equation}\label{eq:Ediss}
    E_{diss}= E_{switch} + E_{excess}
\end{equation}

\noindent Substituting Eqs. (\ref{eq:Edissswitch}), (\ref{eq:Excess_ss}), (\ref{eq:ExpEend}) and (\ref{eq:ExpEqEnd}) in the above equation, we obtain  
\begin{eqnarray}\label{eq:Ediss1}
    E_{diss}&=& - \int_{0}^{T_s}\mathrm{Tr}\Big(\mathbb{D}\hat{H}\Big)dt \\ \nonumber
    &+& \mathrm{Tr}\Big(\hat{\rho}(T_s)\hat{H}(Ts)-\hat{\rho}_{ss}(T_s)\hat{H}_{ss}(T_s)\Big).
\end{eqnarray}

\begin{figure}
\centering
\includegraphics[scale=0.6]{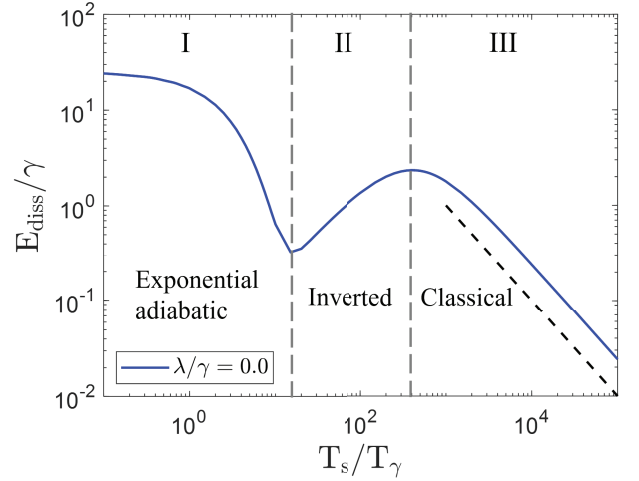}
\caption{Total dissipated energy $E_{diss}$ of an open system as a function of switching time $T_s$ with no reorganization energy ($\lambda$=0). For rapid switching (small $T_s$) in the regime labeled I, $E_{diss}$ exhibits exponential adiabaticity. For intermediate values of $T_s$ in the regime labeled II, we observe the inverted region where $E_{diss}$ actually increases as switching slows down. For large $T_s$ in regime labeled III, $E_{diss}$ follows the classical $1/T_s$ dependence. Here the bias $\Delta$ is varied from $-25\gamma$ to $+25\gamma$ over $T_s$, $T_d=100T_{\gamma}$ and $kT=3\gamma$. The dotted line simply indicates the slope of a  $1/T_s$ dependence on switching time that is characteristic of a classical system.}
\label{fig:TwoDotEdiss0}
\end{figure}

For an open system we solve for the density operator $\hat{\rho}$ using the Lindblad equation in Eq. (\ref{eq:LindbladEquation}). We then calculate the total dissipated energy $E_{diss}$ using Eq. (\ref{eq:Ediss1}). 
$E_{diss}$ is calculated as a function of switching time $T_s$ for different reorganization energies $\lambda$. We first focus on the $\lambda=0$ case shown in Figure \ref{fig:TwoDotEdiss0}, which is our previously reported result. \cite{pidaparthi2021energy} The $E_{diss}$ vs. $T_s$ behavior can be broken into three regimes.

First, the exponentially adiabatic regime at rapid switching (low $T_s$) which is labeled I. When the switching is rapid, there is significant excess energy $E_{excess}$ left in the system at the end of the switching event and the amount of energy dissipated during switching $E_{switch}$ is negligible. The total dissipated energy $E_{diss} \approx E_{excess}$ and decreases exponentially with switching time as shown in Eq. (\ref{eq:TwoStateEexcess}).
 
Second, the inverted regime at intermediate switching speeds labeled II. In this regime, we observe the total dissipated $E_{diss}$ increases as the switching speed decreases. It is surprising that slower switching produces higher dissipation. At these intermediate switching speeds, $E_{excess}$ is negligible due to its exponential dependence on $T_s$. The total dissipated energy is approximately the energy dissipated during switching  $E_{diss} \approx E_{switch}$. When the switching time is much faster than the dissipation time constant $T_d$, $E_{diss}$ is small because the switching is complete before any thermal excitation from the environment can occur. As the switching speed slows down, thermal fluctuations from the environment cause excitations above the ground state. In this regime, the system is moving slowly enough for the excitations to occur, but not slowly enough for them to immediately de-excite and permit the system to closely track the equilibrium occupations. The result is that slowing the switching speed counter-intuitively causes $E_{diss}$ to increase. We have shown that the inverted region does not occur when environmental temperature is zero. The peak of the inverted region occurs at a switching time $T_s \approx 2T_d$. This inverted dependence has been confirmed using a semiclassical treatment.\cite{pidaparthi2021energy}

Third, the classical regime which is labeled III, at slow switching speeds (large $T_s$). Here the total dissipated energy $E_{diss}$ decreases linearly with $T_s$ matching a classical system. At slow switching, the system is close to equilibrium with the environment and any thermal excitation or de-excitation occurs rapidly. But the system is slightly excited and this energy decreases linearly with increasing switching time. The dotted line in Figure \ref{fig:TwoDotEdiss0} showing the $1/T_s$ dependence has been added as a visual guide.

\begin{figure}
\centering
\includegraphics[scale=0.6, trim={0 0 0.2cm 0},clip]{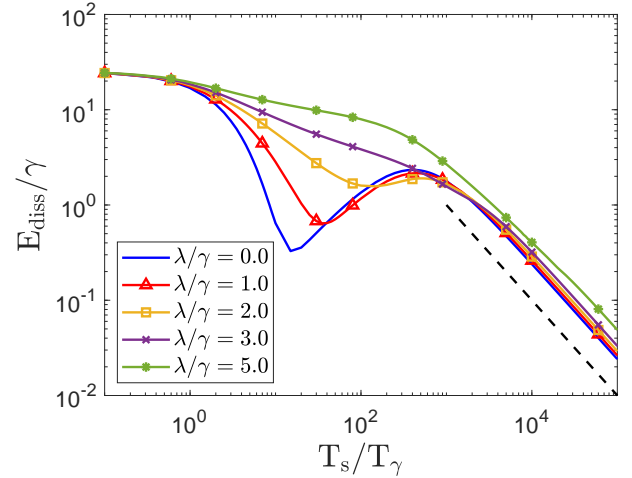}
\caption{Total dissipated energy $E_{diss}$ of an open system as a function of switching time $T_s$ for different reorganization energies $\lambda$. The $\lambda=0$ calculation is the same as shown in Figure \ref{fig:TwoDotEdiss0}. For small values of $\lambda$, $E_{diss}$ is higher in the exponential adiabatic regime. $E_{diss}$ is similar to $E_{diss}$ for $\lambda=0$ in the inverted and classical regimes. For larger values of $\lambda$, $E_{diss}$ is increased in the exponential adiabatic region compared with the $\lambda=0$ case and the inverted region vanishes. At higher $T_s$, $E_{diss}$ shows a classical dependence on $T_s$. Here the bias $\Delta$ is varied from $-25\gamma$ to $+25\gamma$ over $T_s$, $T_d=100T_{\gamma}$ and $kT=3\gamma$. The dotted line is the $1/T_s$ dependence of a classical system. }
\label{fig:TwoDotEdiss}
\end{figure}

The impact of reorganization energy is shown in Figure \ref{fig:TwoDotEdiss}. When the reorganization energy is small ($\lambda/\gamma \lesssim 1$), we observe exponential dependence at rapid switching (small $T_s$). But the rate of $E_{diss}$ decrease is slower. At rapid switching the dissipated energy is approximately equal to the excess energy left at the end of switching event $E_{diss} \approx E_{excess}$. This excess energy increases with increasing reorganization energy as shown in Figure \ref{fig:TwoDotEexcess}. As switching slows down, $E_{diss}$ matches system with no reorganization energy ($\lambda=0$) and we observe the inverted region and the classical dependence.
 
At intermediate reorganization energies ($ 1 \lesssim \lambda/\gamma \lesssim  3 $), the dissipated energy increases in the exponentially adiabatic region. At slower switching $E_{diss}$ meets the inverted dependence at $\lambda=0$. The inverted region gets less pronounced with increasing reorganization energy.  
 
At large  reorganization energies ($\lambda/\gamma \gtrsim 3 $), the dissipated energy is very high at rapid switching. At slower switching, the inverted region completely vanishes and the dissipated energy follows the classical $1/T_s$ dependence. 
 
\begin{figure}
\centering
\includegraphics[scale=0.6]{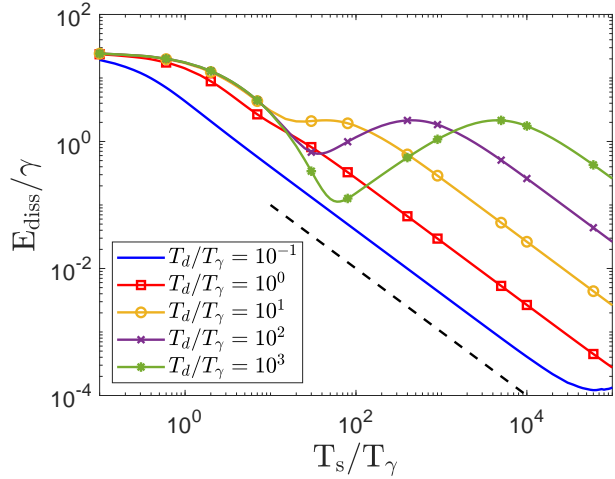}
\caption{Total dissipated energy $E_{diss}$ of an open system as a function of switching time $T_s$ for different values of the dissipation time constant $T_d$ when the reorganization energy $\lambda$ is small. Here the bias $\Delta$ is varied from $-25\gamma$ to $+25\gamma$ over $T_s$, $\lambda=1\gamma$.  and $kT=3\gamma$. The dotted line is the $1/T_s$ dependence of a classical system.}
\label{fig:EdissTd2}
\end{figure}

\begin{figure}
\centering
\includegraphics[scale=0.6]{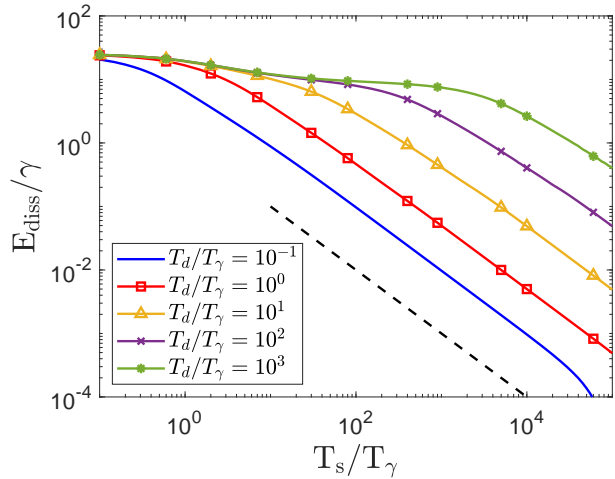}
\caption{Total dissipated energy $E_{diss}$ of an open system as a function of switching time $T_s$ for different values of the dissipation time constant $T_d$ when the reorganization energy $\lambda$ is large. Here the bias $\Delta$ is varied from $-25\gamma$ to $+25\gamma$ over $T_s$, $\lambda=6\gamma$.  and $kT=3\gamma$. Dotted line is the $1/T_s$ dependence of a classical system.}
\label{fig:EdissTd3}
\end{figure}

\begin{figure}
\centering
\includegraphics[scale=0.6]{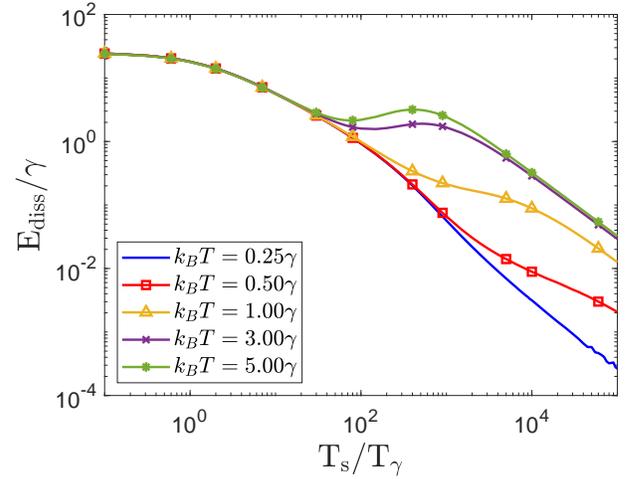}
\caption{Total dissipated energy $E_{diss}$ of an open system as a function of switching time $T_s$ for different values of the environmental temperature $T$. $E_{diss}$ increases as T is increased. The inverted region becomes more pronounced with increasing T. Here the bias $\Delta$ is varied from $-25\gamma$ to $+25\gamma$ over $T_s$, $\lambda=2\gamma$ and $T_d=100T_{\gamma}$. }
\label{fig:EdisskT}
\end{figure}

Figures \ref{fig:EdissTd2} and \ref{fig:EdissTd3} show the impact of environmental coupling when reorganization energy $\lambda=1\gamma$ and $\lambda=6\gamma$ respectively. $T_d/T_{\gamma}$ is varied by four orders of magnitude from $10^{-1}$ to $10^3$ at environmental temperature $k_BT=3\gamma$.  The dotted line shows the $1/T_s$ dependence of a classical system. When the system is very strongly coupled to the environment ($T_d$ is small) any excess energy generated during switching is immediately dissipated and the system acts classically. The system does not exhibit the characteristic quantum adiabatic exponential decrease even at fast switching which is evident in the  $T_d/T_\gamma=0.1$ case which shows  $E_{diss}$ linearly decreasing with $T_s$. For small reorganization energy  $\lambda=1\gamma$ as the coupling becomes weaker, $E_{diss}$ follows the exponential decrease of the isolated system at lower $T_s$, but eventually, as  $T_s$ increases, the  interaction moves dissipation into the classical viscous regime. The inverted region between the two regimes is readily apparent. For larger reorganization energy  $\lambda=6\gamma$, $E_{diss}$ decreases linearly with $T_s$ and increases as environmental coupling weakens.\\
Figure \ref{fig:EdisskT} shows the results for different environmental temperatures  with $T_d=100T_{\gamma}$.  We observe that when switching is rapid, $E_{diss}$ matches the excess the energy of an isolated system. When the temperature is low there is no significant thermal excitation from the environment and $E_{diss}$ decreases monotonically with $T_s$. As the temperature increases $E_{diss}$ starts exhibiting the inverted regime at larger switching times. \\
\section{Discussion}


Molecular relaxation and subsequent lowering of dot energy upon charge occupation has been reported in candidate QCA molecules. We have here examined  the impact of this molecular reorganization energy on the localization and energy dissipation of a two-state QCA cell. 

The Hamiltonian of a QCA cell with reorganization energy, Eq. (\ref{eq:Hamiltonian}), is non-linear due to the effect of ligand distortion. The steady state Hamiltonian depends on the steady state density operator and for small applied biases can have multiple solutions for the same parameters, as seen in Figure \ref{fig:Random}.  We solved for the switching dynamics using the Lindblad equation (\ref{eq:LindbladEquation}) to model the environmental interaction at finite temperature. Molecular QCA cells with non-zero  reorganization energy display hysteresis in the polarization response, as shown in Figure \ref{fig:Polarization4}.

A stored memory bit must necessarily be in a long-lived metastable state that depends on the system's past. Because molecules cannot be addressed individually, the usual QCA approach is to hold binary information not in a single cell, but rather in moving bit packets comprising several cells that are shepherded smoothly around a circuit by the clocking fields.\cite{Lent2003, clocked_molecular_QCA2003} The device feature size can be much smaller than the electrodes that provide the clocking field which move information in computational waves around the circuit architecture. \cite{LentBlair2011, KoggeNetworks2016} The stability of the bit is then due to the large kinetic barrier to flipping all the cells in a bit packet, a stability enhanced by the quantum decoherence caused by the environment.\cite{Blair2013}

The reorganization energy $\lambda$ stabilizes the localized charge when the bias is removed, allowing a single molecular two-state QCA cell to act as a memory cell for storing a single bit. The molecular reorganization energy provides another source of bit stability, namely the kinetic barrier associated with rearranging the atoms within the molecule in addition to moving the electron from site to site. The reorganization energy can be  larger or smaller than $k_BT$, depending on the details of the molecule and its connection to the surrounding environment.

The enhanced bit stability due to the intrinsic molecular reorganization energy comes at the expense of somewhat higher power dissipation for a switching operation. As we have seen in Figure \ref{fig:OnsiteEnergy}, the movement of the dot ligands in response to the electron transfer accelerates the electron transfer process, in a way reminiscent of feedback. The result is lower adiabaticity and more  energy dissipation for the same switching speed.   It is to be noted that this effect is more pronounced at slower switching speeds than at the higher speeds. If the switching speed is faster than the speed at which the ligands can respond, then the reorganization is irrelevant. For small reorganization energies, we observe three distinct regimes in dissipated energy dependence on switching time: an exponential adiabatic ($e^{-aT_s}$) regime, an inverted regime in which dissipation increases with slower switching, and a classical $1/T_s$ dependence for very slow switching. For large reorganization energy, the power dissipation is primarily classical.

From a practical molecular QCA design perspective, one has several key parameters to consider.  The tunneling energy $\gamma$ and therefore $T_{\gamma}$ is determined by the choice  of the linker between the charge centers. The coupling strength to the environment $T_d$ depends on  the precise way the molecule is bonded  to  substrate---essentially the amount of heat-sinking.  The  switching signal $E_c$ applied through electrodes and switching time $T_s$ can be modified through the external driver circuitry. A mixed-valence molecule in a polar solvent typically has a large reorganization energy because many surrounding molecules must move to respond to an electron transfer event. For a surface-bound molecule not in solution, this energy can be much less. We have explored a range of $\lambda$ in our calculations. 
 
The results show a trade-off between enhanced QCA cell bi-stability and energy dissipation.   One might expect that strong coupling to the environment, i.e., strong heat sinking (low $T_d$), would be optimal, but our results show otherwise.  Systems with modest coupling to the substrate  can take advantage of exponential adiabaticity at fast switching speeds (small $T_s$) and therefore optimize net power dissipation.

\section*{Data Availability}
Data sharing is not applicable to this article as no new data were created or analyzed in this study.

\appendix
\nocite{}

%

\end{document}